\documentclass[aps,prl,preprint,superscriptaddress]{revtex4-1}
\usepackage{graphicx}
\usepackage{color}
\usepackage{bm}
\usepackage{braket}
\usepackage{amsmath}

\begin{document}


\title{Anomalous Hall Effect in Kagome Ferrimagnet GdMn$_6$Sn$_6$}
\author{T. Asaba$^{1}$, S. M. Thomas$^{1}$, M. Curtis$^{1}$, J. D. Thompson$^{1}$, E. D. Bauer$^1$, F. Ronning}

\affiliation{
Los Alamos National Laboratory, NM, 87545 USA \\}
\date{\today}
\begin{abstract}
We present magnetotransport data on the ferrimagnet GdMn$_6$Sn$_6$. From the temperature dependent data we are able to extract a large instrinsic contribution to the anomalous Hall effect $\sigma_{xz}^{int} \sim$  32 $\Omega^{-1}cm^{-1}$ and $\sigma_{xy}^{int} \sim$  223 $\Omega^{-1}cm^{-1}$, which is comparable to values found in other systems also containing kagome nets of transition metals. From our transport anisotropy, as well as our density functional theory calculations, we argue that the system is electronically best described as a three dimensional system. Thus, we show that reduced dimensionality is not a strong requirement for obtaining large Berry phase contributions to transport properties. In addition, the coexistence of rare-earth and transition metal magnetism makes the hexagonal MgFe$_6$Ge$_6$ structure type a promising system to tune the electronic and magnetic properties in future studies. 
\end{abstract}

\maketitle

{\bf Introduction} 
The 2D kagome lattice is the marquee system of magnetic frustration \cite{HeltonPRL2007, JoPRL2012, BalentsSpinLiquidNature2010}. This frustration can in principle lead to non-collinear magnetic structures and even novel spin-liquid states. The 2D kagome lattice has also served as the prime model system to understand how strong Berry curvature can lead to a large intrinsic anomalous Hall conductivity (AHC) in non-collinear magnets \cite{ChenPRL2014Mn3Ir, Ohgushi2000, TangPRL2011, ZhuPRL2016}. Motivated by this body of research, experimental studies have revealed large intrinsic contributions to the AHC in layered kagome lattice systems with both antiferromagnetic and ferromagnetic ordering, such as Mn$_3$Sn, Co$_3$Sn$_2$S$_2$, and others \cite{nakatsuji2015large,Kida2011,ye2018massive,nayak2016large,liu2018giant,hirschberger2018skyrmion,U3Ru4Al12}. Perhaps, for ease of comparison with theoretical calculations, it is sometimes argued that the quasi-2D nature of the electronic structure is responsible for the large magnitude of the intrinsic AHC, even though the large ordering temperature is necessarily a reflection of the interlayer magnetic coupling. 

\begin{figure}[h]
	\includegraphics[width=\linewidth]{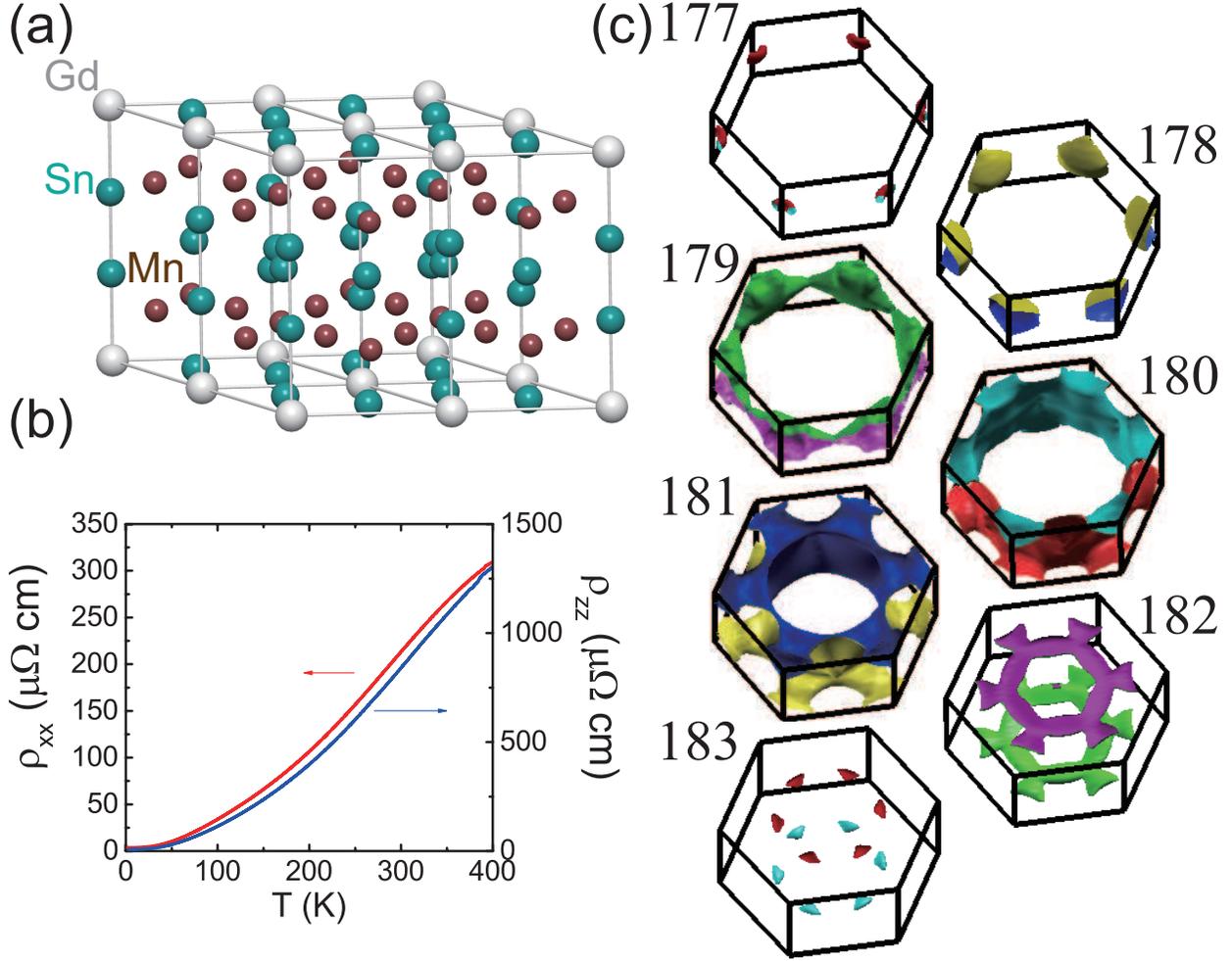}
	\caption{\label{figbasic} (a) Crystal structure of GdMn$_6$Sn$_6$.  (b) In-plane (red) and out-of-plane (blue) resistivity versus temperature. (c) Individual plots of the seven Fermi surface sheets found by DFT calculations in the magnetically ordered state. 
			}
\end{figure}

Here we illustrate that the layered kagome lattice system GdMn$_6$Sn$_6$ also possesses a large intrinsic AHC, although the electronic structure is three dimensional. Fig. \ref{figbasic} (a) shows the crystal structure of GdMn$_6$Sn$_6$. It has a hexagonal structure with space group $P$6/$mmm$ (No. 191), which consists of kagome layers of Mn atoms sandwiched by Sn and Gd atoms. The lattice parameters a and c are 0.552 and 0.902 nm, respectively. The system is a collinear ferrimagnet with $T_c \sim$ 440 K \cite{malaman1999magnetic}. The magnetic moments lie in the ab-plane (easy plane), with collinear but antiparallel Mn and Gd moments.

{\bf Methods} 
Single crystals of GdMn$_6$Sn$_6$ were grown from Sn flux, described in Ref. \cite{Gorbunov2012Gd166}.  The starting elements were placed in an alumina crucible in the ratio Gd:Mn:Sn=1:6:20 and sealed in an evacuated silica ampoule.  The ampoule was heated to 1100 $^{\circ}$C, held at 1100 $^{\circ}$C for 6 hrs., then cooled slowly at 3 $^{\circ}$C/hr. to 550 $^{\circ}$C, at which point the excess Sn flux was removed using a centrifuge.  The orientation of the resulting hexagonal GdMn$_6$Sn$_6$ plates was determined via Laue diffraction. Spin-polarized density functional theory (DFT) calculations of the electronic structure were performed with the WIEN2K code using the PBE functional under the generalized gradient approximation and spin-orbit coupling included via a second variational step \cite{WIEN2K, PBE}.  The number of basis functions were set using a value of $RKmax$ = 8 and a $k$-mesh of 4,000 points in the Brillouin zone.

{\bf Results} 
As shown in the temperature dependence of the resistivity in Fig. \ref{figbasic} (b), both $\rho_{xx}$ and $\rho_{zz}$ have high residual resistivity ratios (RRR) R(300 K)/R(2K) of more than 60, reflecting good metallic behavior. The resistivity anisotropy $\rho_{zz}$/$\rho_{xx}$ stays within a range of 2.2 to 4.2, suggesting that the electronic properties are more 3D-like than 2D-like. This observation is further consistent with the electronic structure calculations. The Fermi surfaces obtained by the DFT calculations are shown in Fig. \ref{figbasic} (c). In the calculation, a ferrimagnetic solution is found with 7.1 $\mu_{B}$ Gd moments antialligned with 2.4 $\mu_{B}$ Mn moments. Within the DFT calculation the size of the atomic moments are defined as the spin density within the muffin-tin radii, of 1.32 $\AA$ for Gd and 1.31 $\AA$ for Mn. The remaining spin polarization that occurs within the Sn muffin-tin radii as well as in the interstitial space of the unit cell is 1.2 $\mu_{B}$ yielding a total moment 6.1 $\mu_{B}/f.u$. in perfect agreement with the experimental result of 6.1 $\mu_{B}/f.u.$ \cite{Gorbunov2012Gd166}. There are seven bands (No. 177-183) crossing the Fermi level. Among them five Fermi surfaces (177-179, 182-183) are 3D-like while only two of them (180-181) have some quasi-2D character. Combining the small resistivity anisotropy with the Fermi surface calculation, in spite of 2D-kagome nets, the electrical properties of the system are dominantly governed by a 3D electronic structure.

\begin{figure}[h]
	\includegraphics[width=\linewidth]{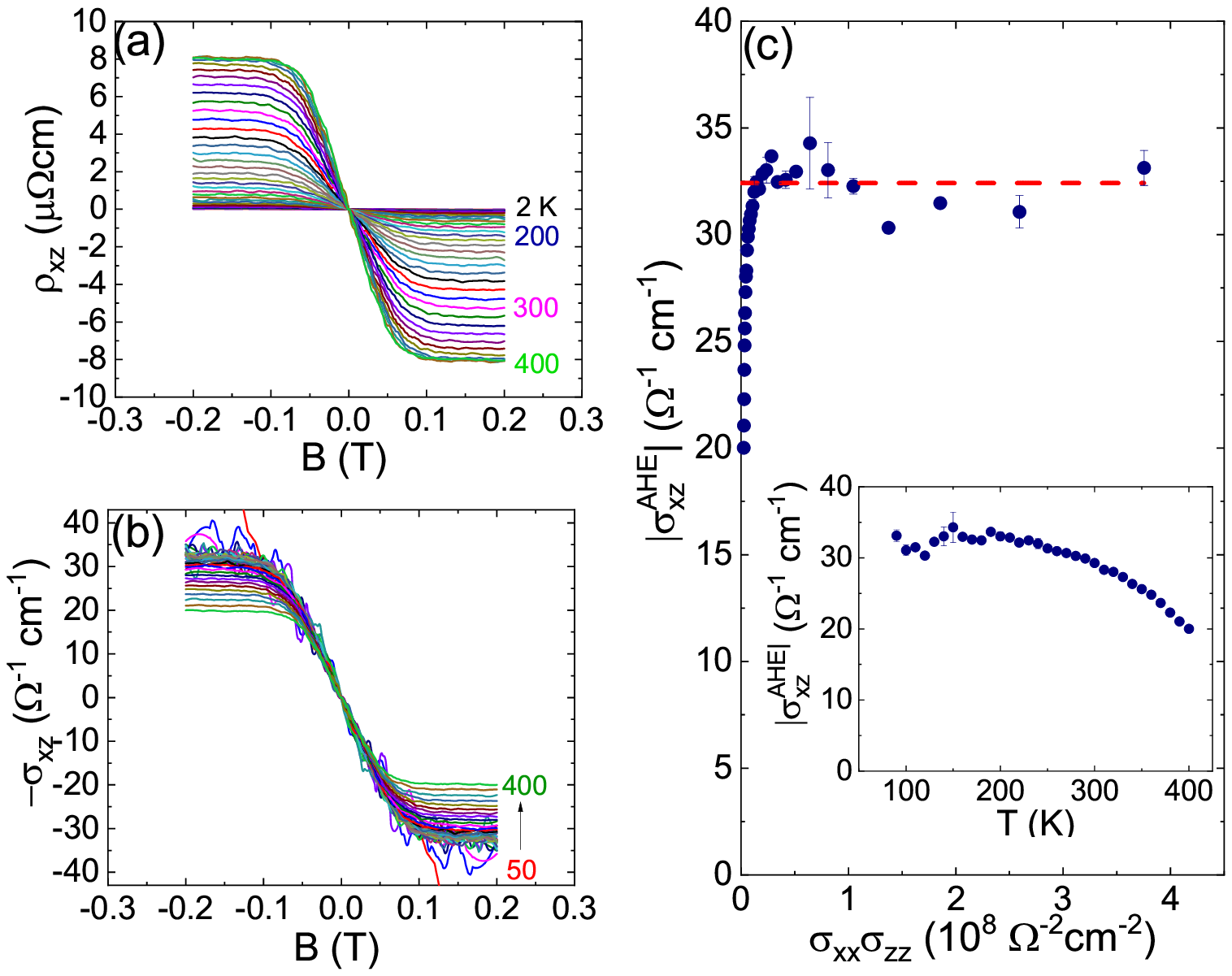}
	\caption{\label{figRxz} 
		{ (a) Transverse resistivity of GdMn$_6$Sn$_6$ versus the applied field along the y-axis for various temperatures. (b) Transverse conductivity versus the applied field along the y-axis for various temperatures. (c) The anomalous Hall conductivity extracted at $B$ = 0.2 T versus the longitudinal conductivity using temperature as an implicit parameter. The dashed line in the limit of zero conductivity provides an estimate of the intrinsic contribution as described in the text. The same data versus temperature is shown in the inset.  The error bars represent the standard error of $\sigma_{xz}^{AHE}$ data between $B$ = 0.15 T and 0.2 T.}
	}

\end{figure}

To evaluate the Berry curvature in this 3D layered kagome system, we first focus on the transverse resistivity response $\rho_{xz}$ with the magnetic field applied parallel to the y-axis  (easy plane). As shown in Fig. \ref{figRxz} (a), at high temperatures, $\rho_{xz}$ is quickly saturated with small fields ($B$ $<$ 0.2 T), reflecting the easy-plane magnetization behavior  shown in the previous study \cite{Gorbunov2012Gd166}, consistent with our own magnetization data (not shown). At low temperatures ($T$ $<$ 50 K), on the other hand, the term that is proportional to the magnetization is hardly seen and $\rho_{xz}$ is dominated by the ordinary Hall component. To further elucidate the intrinsic contribution in $\rho_{xz}$, we calculate the anomalous Hall conductivity as shown in Fig. \ref{figRxz} (b). Interestingly, the magnitude of the step at around zero field in $\sigma_{xz}$ converges to one value above $T$ = 50 K. This fact indicates that the anomalous Hall conductivity is dominated by the intrinsic contribution in this temperature regime. Since the intrinsic AHC is independent of the electron scattering rate and the extrinsic component is not, the contributions to  $\sigma_{xz}=\frac{-\rho_{xz}}{\rho_{xx}\rho_{zz}+\rho_{xz}^2}$ can be described as
\begin{equation} \label{AHEeq}
\sigma_{xz} = \sigma_{xz}^{N} + \sigma_{xz}^{ext} + \sigma_{xz}^{int}
\end{equation}
where $\sigma_{xz}^{N}$ is the ordinary Hall conductivity, $\sigma_{xz}^{ext}$ is an anomalous component, which can arise from skew scattering or the side-jump mechanism, and $\sigma_{xz}^{int}$ is an intrinsic component arising from the Berry curvature of the electronic structure. The first two terms depend strongly on the scattering rate, while the latter is independent of it. From the resistivity shown in Fig. \ref{figbasic} (b) we can infer that the scattering rate is highly temperature dependent. Consequently, the lack of temperature dependence in $\sigma_{xz}$ above 50 K, reflects the dominance of the intrinsic Berry curvature derived $\sigma_{xz}^{int}$ contribution.

To extract the amplitude of the intrinsic component, the  AHC versus the longitudinal conductivity at $B$ = 0.2 T is shown in Fig. \ref{figRxz} (c). Since $B$ = 0.2 T is high enough to saturate the anomalous Hall component but low enough for the ordinary Hall component to be negligible, we can assume that the transverse response is dominated by the anomalous Hall contribution. To separate the intrinsic contribution to the AHC from the extrinsic one, we use the equation:

\begin{equation} \label{AHEeq2}
\sigma_{xz}^{AHE} = \sigma_{xz}^{int} + \alpha\sigma_{xx} \sigma_{zz}
\end{equation}
where $\alpha$ represents the extrinsic component, which is scattering-rate and thereby conductivity-dependent \cite{tian2009proper,ye2018massive}.

From eq. \ref{AHEeq2}, $\sigma_{xz}^{int}$ can be obtained as $\sigma_{xx}\sigma_{zz} \rightarrow 0$ at moderately high temperatures. The down turn of $\sigma_{xz}$ at the highest temperatures is a consequence of the vanishing Berry curvature associated with suppressing the ferromagnetic component of the magnetization to zero as the Curie temperature is approached, and should be ignored.  In our case, both $\sigma_{xz}$ and $\sigma_{xy}$ (shown below) are approximately temperature-independent. Thus we set $\alpha$=0 and fit the high temperature data of the hall conductivity to a constant (red dashed line in Fig.\ref{figRxz}c).  The resulting  $\sigma_{xz}^{int} =$  32 $\pm$ 2 $\Omega^{-1}$cm$^{-1}$ corresponds to 0.073 $e^2/ha$, where $e$, $h$ and $a$ are electron charge, the Plank constant, and the lattice parameter, respectively. 


\begin{figure}[h]
	\includegraphics[width=\linewidth]{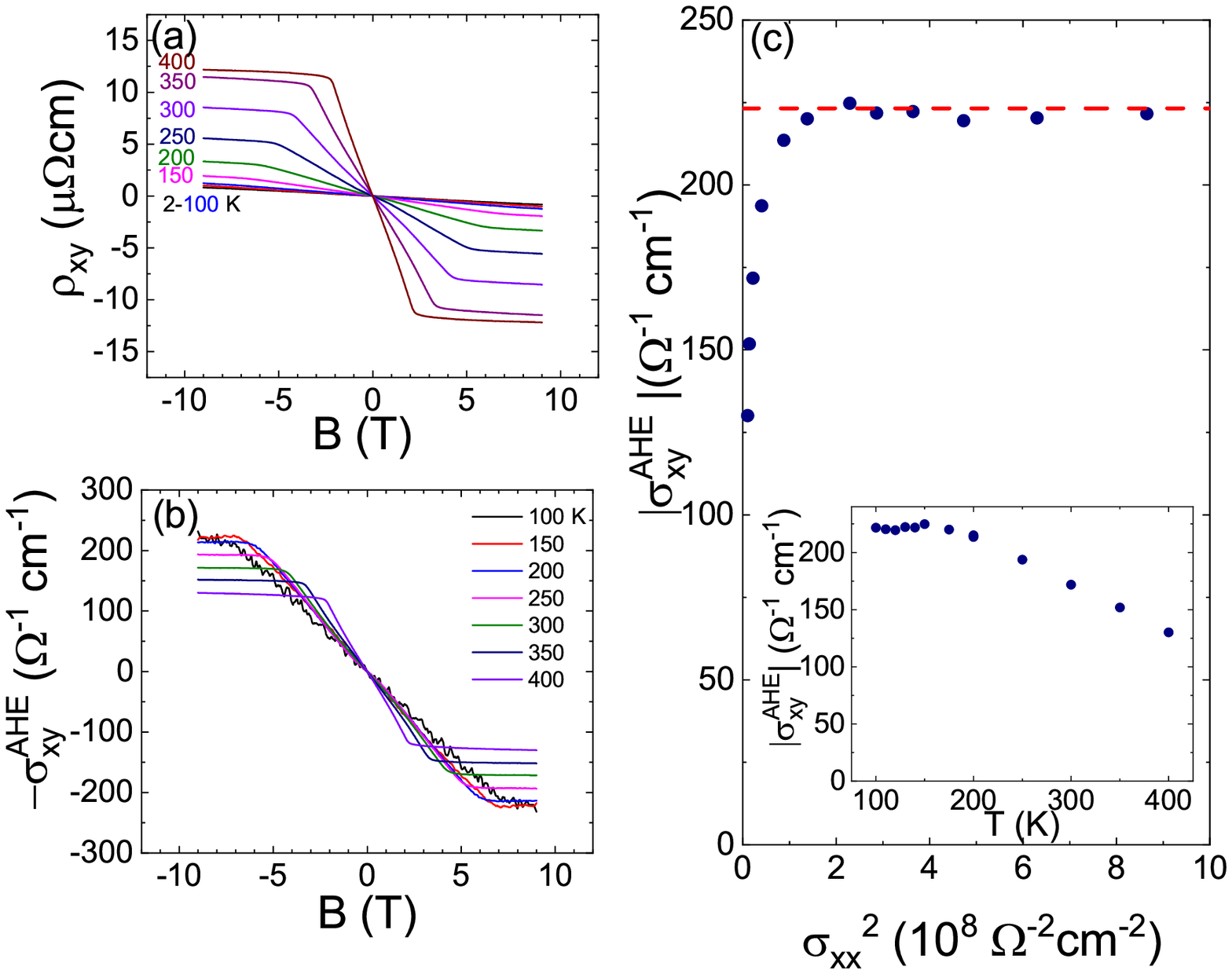}
	\caption{\label{figRxy} 
		{(a)-(c) Same as in Fig. \ref{figRxz} with field applied along the z-axis. For (b), the ordinary Hall contribution is subtracted, as described in the main text.
	}
}
\end{figure}

Next, the Hall response with the magnetic field applied parallel to the $c$-axis (hard axis) is shown in Fig. \ref{figRxy} (a). At low temperatures $\rho_{xy}$ is almost linear with the magnetic field up to $B$ = 6 T, at which point the magnetization becomes saturated. This behavior is consistent with the magnetization curve\cite{Gorbunov2012Gd166}. Above the saturation field, the slope of the Hall signal is almost constant, reflecting the ordinary Hall signal. Since the saturation field is much higher than that with $B\parallel a$, the ordinary Hall signal is no longer negligible and it is necessary to account for it to obtain the anomalous Hall contribution. At $T$ = 50 K we estimate the ordinary Hall coefficient $R_H$ = 0.11 $\mu\Omega$cm/T, which corresponds to a carrier density $n$ of 5.7 $\times$ $10^{21}$ /cm$^3$.  We then obtain the anomalous Hall component by subtracting the ordinary  Hall component. The resulting $\sigma_{xy}^{AHE}$ is shown in Fig. \ref{figRxy} (b). Similar to $\sigma_{xz}^{AHE}$, the saturated value of $\sigma_{xy}^{AHE}$ is almost constant, indicating that the intrinsic contribution dominates the extrinsic anomalous Hall effect. With the same analysis performed for $\sigma_{xz}$ above, we estimate the intrinsic anomalous Hall component  $\sigma_{xy}^{int} \sim$  223 $\pm$ 2 $\Omega^{-1}$cm$^{-1}$. This corresponds to  0.54 $e^2/hc$, where $c$ is the lattice constant. Since a unit cell contains two kagome layers, this contribution is reduced to 0.27  $e^2/h$ per single kagome layer.



{\bf Discussion}

A 2D-electron gas can possess a large and quantized Hall response in units of e$^2$/$h$ \cite{QHE}. Although no longer quantized, one may anticipate that a 3D material may possess an enhanced Berry curvature contribution to the Hall conductivity in systems where the electronic structure remains quasi 2D. Here, we have demonstrated that GdMn$_6$Sn$_6$ possesses a large intrinsic contribution to the Hall conductivity. In addition, both Fermi surface calculations and resistivity anisotropy indicate the three-dimensional electronic structure in this system. The observed intrinsic AHC of 223 $\Omega^{-1}$ cm$^{-1}$ for GdMn$_6$Sn$_6$, however, is comparable to that for Fe$_3$Sn$_2$, which is attributed to reduced dimensionality \cite{ye2018massive}. Note, pure elemental iron with a cubic crystal structure also shows a large intrinsic AHC greater than 1000 $\Omega^{-1}$ cm$^{-1}$ \cite{miyasato2007crossover}. This illustrates that a quasi-2D electronic structure is not a prerequisite for generating large Berry curvature effects that can be easily manipulated, for instance, by small magnetic fields. 


GdMn$_6$Sn$_6$ is just one member of a large family of RT$_6$X$_6$, where R is a rare-earth, T is a transition metal, and X is typically either Sn or Ge. For RMn$_6$Sn$_6$ compounds, the system orders with an antiferromagnetic spiral for light  or heavy rare-earths (Sc,Y,Eu,Lu), and  for moderate rare-earths (Gd-Ho), a ferrimagnetic ground state is found \cite{venturini1991magnetic,venturini1996incommensurate,malaman1999magnetic}.  Interestingly, for YMn$_6$Sn$_6$ a field polarized state could be achieved at $\sim$ 13 T, which is similar to our case where the Mn moments are fully polarized \cite{wang2019near}. The AHC found in YMn$_6$Sn$_6$ is $\sim$ 45 $\Omega^{-1}$ cm$^{-1}$ and 300 $\Omega^{-1}$ cm$^{-1}$ for $\sigma_{xy}$ and $\sigma_{xz}$, respectively \cite{wang2019near}. This is very comparable in magnitude to the values we find in GdMn$_6$Sn$_6$, but with the reverse anisotropy.  The difference between these two compounds indicates that the Berry curvature of RMn$_6$Sn$_6$ could be easily tuned by the rare-earth elements. Therefore, this suggests that the RT$_6$X$_6$ system in general should be an excellent system to learn how to manipulate the Berry curvature via the tuning knob provided by the localized magnetism on the rare-earth site.

{\bf Acknowledgement} 
This work was carried out under the auspices of the U.S. Department of Energy, Office of Science, Basic Energy Sciences, Materials Sciences and Engineering Division. TA acknowledges support of the LANL Directors Postdoctoral Funding LDRD program.

\bibliography{RefGdMn6Sn6}

\begin{thebibliography}{24}%
\makeatletter
\providecommand \@ifxundefined [1]{%
 \@ifx{#1\undefined}
}%
\providecommand \@ifnum [1]{%
 \ifnum #1\expandafter \@firstoftwo
 \else \expandafter \@secondoftwo
 \fi
}%
\providecommand \@ifx [1]{%
 \ifx #1\expandafter \@firstoftwo
 \else \expandafter \@secondoftwo
 \fi
}%
\providecommand \natexlab [1]{#1}%
\providecommand \enquote  [1]{``#1''}%
\providecommand \bibnamefont  [1]{#1}%
\providecommand \bibfnamefont [1]{#1}%
\providecommand \citenamefont [1]{#1}%
\providecommand \href@noop [0]{\@secondoftwo}%
\providecommand \href [0]{\begingroup \@sanitize@url \@href}%
\providecommand \@href[1]{\@@startlink{#1}\@@href}%
\providecommand \@@href[1]{\endgroup#1\@@endlink}%
\providecommand \@sanitize@url [0]{\catcode `\\12\catcode `\$12\catcode
  `\&12\catcode `\#12\catcode `\^12\catcode `\_12\catcode `\%12\relax}%
\providecommand \@@startlink[1]{}%
\providecommand \@@endlink[0]{}%
\providecommand \url  [0]{\begingroup\@sanitize@url \@url }%
\providecommand \@url [1]{\endgroup\@href {#1}{\urlprefix }}%
\providecommand \urlprefix  [0]{URL }%
\providecommand \Eprint [0]{\href }%
\providecommand \doibase [0]{http://dx.doi.org/}%
\providecommand \selectlanguage [0]{\@gobble}%
\providecommand \bibinfo  [0]{\@secondoftwo}%
\providecommand \bibfield  [0]{\@secondoftwo}%
\providecommand \translation [1]{[#1]}%
\providecommand \BibitemOpen [0]{}%
\providecommand \bibitemStop [0]{}%
\providecommand \bibitemNoStop [0]{.\EOS\space}%
\providecommand \EOS [0]{\spacefactor3000\relax}%
\providecommand \BibitemShut  [1]{\csname bibitem#1\endcsname}%
\let\auto@bib@innerbib\@empty
\bibitem [{\citenamefont {Helton}\ \emph {et~al.}(2007)\citenamefont {Helton},
  \citenamefont {Matan}, \citenamefont {Shores}, \citenamefont {Nytko},
  \citenamefont {Bartlett}, \citenamefont {Yoshida}, \citenamefont {Takano},
  \citenamefont {Suslov}, \citenamefont {Qiu}, \citenamefont {Chung},
  \citenamefont {Nocera},\ and\ \citenamefont {Lee}}]{HeltonPRL2007}%
  \BibitemOpen
  \bibfield  {author} {\bibinfo {author} {\bibfnamefont {J.~S.}\ \bibnamefont
  {Helton}}, \bibinfo {author} {\bibfnamefont {K.}~\bibnamefont {Matan}},
  \bibinfo {author} {\bibfnamefont {M.~P.}\ \bibnamefont {Shores}}, \bibinfo
  {author} {\bibfnamefont {E.~A.}\ \bibnamefont {Nytko}}, \bibinfo {author}
  {\bibfnamefont {B.~M.}\ \bibnamefont {Bartlett}}, \bibinfo {author}
  {\bibfnamefont {Y.}~\bibnamefont {Yoshida}}, \bibinfo {author} {\bibfnamefont
  {Y.}~\bibnamefont {Takano}}, \bibinfo {author} {\bibfnamefont
  {A.}~\bibnamefont {Suslov}}, \bibinfo {author} {\bibfnamefont
  {Y.}~\bibnamefont {Qiu}}, \bibinfo {author} {\bibfnamefont {J.-H.}\
  \bibnamefont {Chung}}, \bibinfo {author} {\bibfnamefont {D.~G.}\ \bibnamefont
  {Nocera}}, \ and\ \bibinfo {author} {\bibfnamefont {Y.~S.}\ \bibnamefont
  {Lee}},\ }\href {\doibase 10.1103/PhysRevLett.98.107204} {\bibfield
  {journal} {\bibinfo  {journal} {Phys. Rev. Lett.}\ }\textbf {\bibinfo
  {volume} {98}},\ \bibinfo {pages} {107204} (\bibinfo {year}
  {2007})}\BibitemShut {NoStop}%
\bibitem [{\citenamefont {Jo}\ \emph {et~al.}(2012)\citenamefont {Jo},
  \citenamefont {Guzman}, \citenamefont {Thomas}, \citenamefont {Hosur},
  \citenamefont {Vishwanath},\ and\ \citenamefont {Stamper-Kurn}}]{JoPRL2012}%
  \BibitemOpen
  \bibfield  {author} {\bibinfo {author} {\bibfnamefont {G.-B.}\ \bibnamefont
  {Jo}}, \bibinfo {author} {\bibfnamefont {J.}~\bibnamefont {Guzman}}, \bibinfo
  {author} {\bibfnamefont {C.~K.}\ \bibnamefont {Thomas}}, \bibinfo {author}
  {\bibfnamefont {P.}~\bibnamefont {Hosur}}, \bibinfo {author} {\bibfnamefont
  {A.}~\bibnamefont {Vishwanath}}, \ and\ \bibinfo {author} {\bibfnamefont
  {D.~M.}\ \bibnamefont {Stamper-Kurn}},\ }\href {\doibase
  10.1103/PhysRevLett.108.045305} {\bibfield  {journal} {\bibinfo  {journal}
  {Phys. Rev. Lett.}\ }\textbf {\bibinfo {volume} {108}},\ \bibinfo {pages}
  {045305} (\bibinfo {year} {2012})}\BibitemShut {NoStop}%
\bibitem [{\citenamefont {Balents}(2010)}]{BalentsSpinLiquidNature2010}%
  \BibitemOpen
  \bibfield  {author} {\bibinfo {author} {\bibfnamefont {L.}~\bibnamefont
  {Balents}},\ }\href@noop {} {\bibfield  {journal} {\bibinfo  {journal}
  {Nature}\ }\textbf {\bibinfo {volume} {464}},\ \bibinfo {pages} {199–208}
  (\bibinfo {year} {2010})}\BibitemShut {NoStop}%
\bibitem [{\citenamefont {Chen}\ \emph {et~al.}(2014)\citenamefont {Chen},
  \citenamefont {Niu},\ and\ \citenamefont {MacDonald}}]{ChenPRL2014Mn3Ir}%
  \BibitemOpen
  \bibfield  {author} {\bibinfo {author} {\bibfnamefont {H.}~\bibnamefont
  {Chen}}, \bibinfo {author} {\bibfnamefont {Q.}~\bibnamefont {Niu}}, \ and\
  \bibinfo {author} {\bibfnamefont {A.~H.}\ \bibnamefont {MacDonald}},\ }\href
  {\doibase 10.1103/PhysRevLett.112.017205} {\bibfield  {journal} {\bibinfo
  {journal} {Phys. Rev. Lett.}\ }\textbf {\bibinfo {volume} {112}},\ \bibinfo
  {pages} {017205} (\bibinfo {year} {2014})}\BibitemShut {NoStop}%
\bibitem [{\citenamefont {Ohgushi}\ \emph {et~al.}(2000)\citenamefont
  {Ohgushi}, \citenamefont {Murakami},\ and\ \citenamefont
  {Nagaosa}}]{Ohgushi2000}%
  \BibitemOpen
  \bibfield  {author} {\bibinfo {author} {\bibfnamefont {K.}~\bibnamefont
  {Ohgushi}}, \bibinfo {author} {\bibfnamefont {S.}~\bibnamefont {Murakami}}, \
  and\ \bibinfo {author} {\bibfnamefont {N.}~\bibnamefont {Nagaosa}},\ }\href
  {\doibase 10.1103/PhysRevB.62.R6065} {\bibfield  {journal} {\bibinfo
  {journal} {Phys. Rev. B}\ }\textbf {\bibinfo {volume} {62}},\ \bibinfo
  {pages} {R6065} (\bibinfo {year} {2000})}\BibitemShut {NoStop}%
\bibitem [{\citenamefont {Tang}\ \emph {et~al.}(2011)\citenamefont {Tang},
  \citenamefont {Mei},\ and\ \citenamefont {Wen}}]{TangPRL2011}%
  \BibitemOpen
  \bibfield  {author} {\bibinfo {author} {\bibfnamefont {E.}~\bibnamefont
  {Tang}}, \bibinfo {author} {\bibfnamefont {J.-W.}\ \bibnamefont {Mei}}, \
  and\ \bibinfo {author} {\bibfnamefont {X.-G.}\ \bibnamefont {Wen}},\ }\href
  {\doibase 10.1103/PhysRevLett.106.236802} {\bibfield  {journal} {\bibinfo
  {journal} {Phys. Rev. Lett.}\ }\textbf {\bibinfo {volume} {106}},\ \bibinfo
  {pages} {236802} (\bibinfo {year} {2011})}\BibitemShut {NoStop}%
\bibitem [{\citenamefont {Zhu}\ \emph {et~al.}(2016)\citenamefont {Zhu},
  \citenamefont {Gong}, \citenamefont {Zeng}, \citenamefont {Fu},\ and\
  \citenamefont {Sheng}}]{ZhuPRL2016}%
  \BibitemOpen
  \bibfield  {author} {\bibinfo {author} {\bibfnamefont {W.}~\bibnamefont
  {Zhu}}, \bibinfo {author} {\bibfnamefont {S.-S.}\ \bibnamefont {Gong}},
  \bibinfo {author} {\bibfnamefont {T.-S.}\ \bibnamefont {Zeng}}, \bibinfo
  {author} {\bibfnamefont {L.}~\bibnamefont {Fu}}, \ and\ \bibinfo {author}
  {\bibfnamefont {D.~N.}\ \bibnamefont {Sheng}},\ }\href {\doibase
  10.1103/PhysRevLett.117.096402} {\bibfield  {journal} {\bibinfo  {journal}
  {Phys. Rev. Lett.}\ }\textbf {\bibinfo {volume} {117}},\ \bibinfo {pages}
  {096402} (\bibinfo {year} {2016})}\BibitemShut {NoStop}%
\bibitem [{\citenamefont {Nakatsuji}\ \emph {et~al.}(2015)\citenamefont
  {Nakatsuji}, \citenamefont {Kiyohara},\ and\ \citenamefont
  {Higo}}]{nakatsuji2015large}%
  \BibitemOpen
  \bibfield  {author} {\bibinfo {author} {\bibfnamefont {S.}~\bibnamefont
  {Nakatsuji}}, \bibinfo {author} {\bibfnamefont {N.}~\bibnamefont {Kiyohara}},
  \ and\ \bibinfo {author} {\bibfnamefont {T.}~\bibnamefont {Higo}},\
  }\href@noop {} {\bibfield  {journal} {\bibinfo  {journal} {Nature}\ }\textbf
  {\bibinfo {volume} {527}},\ \bibinfo {pages} {212} (\bibinfo {year}
  {2015})}\BibitemShut {NoStop}%
\bibitem [{\citenamefont {Kida}\ \emph {et~al.}(2011)\citenamefont {Kida},
  \citenamefont {Fenner}, \citenamefont {Dee}, \citenamefont {Terasaki},
  \citenamefont {Hagiwara},\ and\ \citenamefont {Wills}}]{Kida2011}%
  \BibitemOpen
  \bibfield  {author} {\bibinfo {author} {\bibfnamefont {T.}~\bibnamefont
  {Kida}}, \bibinfo {author} {\bibfnamefont {L.~A.}\ \bibnamefont {Fenner}},
  \bibinfo {author} {\bibfnamefont {A.~A.}\ \bibnamefont {Dee}}, \bibinfo
  {author} {\bibfnamefont {I.}~\bibnamefont {Terasaki}}, \bibinfo {author}
  {\bibfnamefont {M.}~\bibnamefont {Hagiwara}}, \ and\ \bibinfo {author}
  {\bibfnamefont {A.~S.}\ \bibnamefont {Wills}},\ }\href {\doibase
  10.1088/0953-8984/23/11/112205} {\bibfield  {journal} {\bibinfo  {journal}
  {Journal of Physics: Condensed Matter}\ }\textbf {\bibinfo {volume} {23}},\
  \bibinfo {pages} {112205} (\bibinfo {year} {2011})}\BibitemShut {NoStop}%
\bibitem [{\citenamefont {Ye}\ \emph {et~al.}(2018)\citenamefont {Ye},
  \citenamefont {Kang}, \citenamefont {Liu}, \citenamefont {Von~Cube},
  \citenamefont {Wicker}, \citenamefont {Suzuki}, \citenamefont {Jozwiak},
  \citenamefont {Bostwick}, \citenamefont {Rotenberg}, \citenamefont {Bell}
  \emph {et~al.}}]{ye2018massive}%
  \BibitemOpen
  \bibfield  {author} {\bibinfo {author} {\bibfnamefont {L.}~\bibnamefont
  {Ye}}, \bibinfo {author} {\bibfnamefont {M.}~\bibnamefont {Kang}}, \bibinfo
  {author} {\bibfnamefont {J.}~\bibnamefont {Liu}}, \bibinfo {author}
  {\bibfnamefont {F.}~\bibnamefont {Von~Cube}}, \bibinfo {author}
  {\bibfnamefont {C.~R.}\ \bibnamefont {Wicker}}, \bibinfo {author}
  {\bibfnamefont {T.}~\bibnamefont {Suzuki}}, \bibinfo {author} {\bibfnamefont
  {C.}~\bibnamefont {Jozwiak}}, \bibinfo {author} {\bibfnamefont
  {A.}~\bibnamefont {Bostwick}}, \bibinfo {author} {\bibfnamefont
  {E.}~\bibnamefont {Rotenberg}}, \bibinfo {author} {\bibfnamefont {D.~C.}\
  \bibnamefont {Bell}},  \emph {et~al.},\ }\href@noop {} {\bibfield  {journal}
  {\bibinfo  {journal} {Nature}\ }\textbf {\bibinfo {volume} {555}},\ \bibinfo
  {pages} {638} (\bibinfo {year} {2018})}\BibitemShut {NoStop}%
\bibitem [{\citenamefont {Nayak}\ \emph {et~al.}(2016)\citenamefont {Nayak},
  \citenamefont {Fischer}, \citenamefont {Sun}, \citenamefont {Yan},
  \citenamefont {Karel}, \citenamefont {Komarek}, \citenamefont {Shekhar},
  \citenamefont {Kumar}, \citenamefont {Schnelle}, \citenamefont {K{\"u}bler}
  \emph {et~al.}}]{nayak2016large}%
  \BibitemOpen
  \bibfield  {author} {\bibinfo {author} {\bibfnamefont {A.~K.}\ \bibnamefont
  {Nayak}}, \bibinfo {author} {\bibfnamefont {J.~E.}\ \bibnamefont {Fischer}},
  \bibinfo {author} {\bibfnamefont {Y.}~\bibnamefont {Sun}}, \bibinfo {author}
  {\bibfnamefont {B.}~\bibnamefont {Yan}}, \bibinfo {author} {\bibfnamefont
  {J.}~\bibnamefont {Karel}}, \bibinfo {author} {\bibfnamefont {A.~C.}\
  \bibnamefont {Komarek}}, \bibinfo {author} {\bibfnamefont {C.}~\bibnamefont
  {Shekhar}}, \bibinfo {author} {\bibfnamefont {N.}~\bibnamefont {Kumar}},
  \bibinfo {author} {\bibfnamefont {W.}~\bibnamefont {Schnelle}}, \bibinfo
  {author} {\bibfnamefont {J.}~\bibnamefont {K{\"u}bler}},  \emph {et~al.},\
  }\href@noop {} {\bibfield  {journal} {\bibinfo  {journal} {Science advances}\
  }\textbf {\bibinfo {volume} {2}},\ \bibinfo {pages} {e1501870} (\bibinfo
  {year} {2016})}\BibitemShut {NoStop}%
\bibitem [{\citenamefont {Liu}\ \emph {et~al.}(2018)\citenamefont {Liu},
  \citenamefont {Sun}, \citenamefont {Kumar}, \citenamefont {Muechler},
  \citenamefont {Sun}, \citenamefont {Jiao}, \citenamefont {Yang},
  \citenamefont {Liu}, \citenamefont {Liang}, \citenamefont {Xu} \emph
  {et~al.}}]{liu2018giant}%
  \BibitemOpen
  \bibfield  {author} {\bibinfo {author} {\bibfnamefont {E.}~\bibnamefont
  {Liu}}, \bibinfo {author} {\bibfnamefont {Y.}~\bibnamefont {Sun}}, \bibinfo
  {author} {\bibfnamefont {N.}~\bibnamefont {Kumar}}, \bibinfo {author}
  {\bibfnamefont {L.}~\bibnamefont {Muechler}}, \bibinfo {author}
  {\bibfnamefont {A.}~\bibnamefont {Sun}}, \bibinfo {author} {\bibfnamefont
  {L.}~\bibnamefont {Jiao}}, \bibinfo {author} {\bibfnamefont {S.-Y.}\
  \bibnamefont {Yang}}, \bibinfo {author} {\bibfnamefont {D.}~\bibnamefont
  {Liu}}, \bibinfo {author} {\bibfnamefont {A.}~\bibnamefont {Liang}}, \bibinfo
  {author} {\bibfnamefont {Q.}~\bibnamefont {Xu}},  \emph {et~al.},\
  }\href@noop {} {\bibfield  {journal} {\bibinfo  {journal} {Nature physics}\
  }\textbf {\bibinfo {volume} {14}},\ \bibinfo {pages} {1125} (\bibinfo {year}
  {2018})}\BibitemShut {NoStop}%
\bibitem [{\citenamefont {Hirschberger}\ \emph {et~al.}(2018)\citenamefont
  {Hirschberger}, \citenamefont {Nakajima}, \citenamefont {Gao}, \citenamefont
  {Peng}, \citenamefont {Kikkawa}, \citenamefont {Kurumaji}, \citenamefont
  {Kriener}, \citenamefont {Yamasaki}, \citenamefont {Sagayama}, \citenamefont
  {Nakao} \emph {et~al.}}]{hirschberger2018skyrmion}%
  \BibitemOpen
  \bibfield  {author} {\bibinfo {author} {\bibfnamefont {M.}~\bibnamefont
  {Hirschberger}}, \bibinfo {author} {\bibfnamefont {T.}~\bibnamefont
  {Nakajima}}, \bibinfo {author} {\bibfnamefont {S.}~\bibnamefont {Gao}},
  \bibinfo {author} {\bibfnamefont {L.}~\bibnamefont {Peng}}, \bibinfo {author}
  {\bibfnamefont {A.}~\bibnamefont {Kikkawa}}, \bibinfo {author} {\bibfnamefont
  {T.}~\bibnamefont {Kurumaji}}, \bibinfo {author} {\bibfnamefont
  {M.}~\bibnamefont {Kriener}}, \bibinfo {author} {\bibfnamefont
  {Y.}~\bibnamefont {Yamasaki}}, \bibinfo {author} {\bibfnamefont
  {H.}~\bibnamefont {Sagayama}}, \bibinfo {author} {\bibfnamefont
  {H.}~\bibnamefont {Nakao}},  \emph {et~al.},\ }\href@noop {} {\bibfield
  {journal} {\bibinfo  {journal} {arXiv preprint arXiv:1812.02553}\ } (\bibinfo
  {year} {2018})}\BibitemShut {NoStop}%
\bibitem [{U3R()}]{U3Ru4Al12}%
  \BibitemOpen
  \href@noop {} {\bibinfo  {journal} {Under review}\ }\BibitemShut {NoStop}%
\bibitem [{\citenamefont {Malaman}\ \emph {et~al.}(1999)\citenamefont
  {Malaman}, \citenamefont {Venturini}, \citenamefont {Welter}, \citenamefont
  {Sanchez}, \citenamefont {Vulliet},\ and\ \citenamefont
  {Ressouche}}]{malaman1999magnetic}%
  \BibitemOpen
\bibfield  {journal} {  }\bibfield  {author} {\bibinfo {author} {\bibfnamefont
  {B.}~\bibnamefont {Malaman}}, \bibinfo {author} {\bibfnamefont
  {G.}~\bibnamefont {Venturini}}, \bibinfo {author} {\bibfnamefont
  {R.}~\bibnamefont {Welter}}, \bibinfo {author} {\bibfnamefont
  {J.}~\bibnamefont {Sanchez}}, \bibinfo {author} {\bibfnamefont
  {P.}~\bibnamefont {Vulliet}}, \ and\ \bibinfo {author} {\bibfnamefont
  {E.}~\bibnamefont {Ressouche}},\ }\href@noop {} {\bibfield  {journal}
  {\bibinfo  {journal} {Journal of magnetism and magnetic materials}\ }\textbf
  {\bibinfo {volume} {202}},\ \bibinfo {pages} {519} (\bibinfo {year}
  {1999})}\BibitemShut {NoStop}%
\bibitem [{\citenamefont {Gorbunov}\ \emph {et~al.}(2012)\citenamefont
  {Gorbunov}, \citenamefont {Kuz’min}, \citenamefont
  {Uhl{\'\i}{\v{r}}ov{\'a}}, \citenamefont {{\v{Z}}{\'a}{\v{c}}ek},
  \citenamefont {Richter}, \citenamefont {Skourski},\ and\ \citenamefont
  {Andreev}}]{Gorbunov2012Gd166}%
  \BibitemOpen
  \bibfield  {author} {\bibinfo {author} {\bibfnamefont {D.}~\bibnamefont
  {Gorbunov}}, \bibinfo {author} {\bibfnamefont {M.}~\bibnamefont {Kuz’min}},
  \bibinfo {author} {\bibfnamefont {K.}~\bibnamefont
  {Uhl{\'\i}{\v{r}}ov{\'a}}}, \bibinfo {author} {\bibfnamefont
  {M.}~\bibnamefont {{\v{Z}}{\'a}{\v{c}}ek}}, \bibinfo {author} {\bibfnamefont
  {M.}~\bibnamefont {Richter}}, \bibinfo {author} {\bibfnamefont
  {Y.}~\bibnamefont {Skourski}}, \ and\ \bibinfo {author} {\bibfnamefont
  {A.}~\bibnamefont {Andreev}},\ }\href@noop {} {\bibfield  {journal} {\bibinfo
   {journal} {Journal of Alloys and Compounds}\ }\textbf {\bibinfo {volume}
  {519}},\ \bibinfo {pages} {47} (\bibinfo {year} {2012})}\BibitemShut
  {NoStop}%
\bibitem [{WIE()}]{WIEN2K}%
  \BibitemOpen
  \href@noop {} {\ }\bibinfo {note} {P. Blaha et al., An Augmented Plane Wave +
  Local Orbitals Program for Calculating Crystal Properties (K. Schwarz, Tech.
  UniversitA at Wien, Austria, 2001).}\BibitemShut {Stop}%
\bibitem [{\citenamefont {Perdew}\ \emph {et~al.}(1996)\citenamefont {Perdew},
  \citenamefont {Burke},\ and\ \citenamefont {Ernzerhof}}]{PBE}%
  \BibitemOpen
  \bibfield  {author} {\bibinfo {author} {\bibfnamefont {J.~P.}\ \bibnamefont
  {Perdew}}, \bibinfo {author} {\bibfnamefont {K.}~\bibnamefont {Burke}}, \
  and\ \bibinfo {author} {\bibfnamefont {M.}~\bibnamefont {Ernzerhof}},\ }\href
  {\doibase 10.1103/PhysRevLett.77.3865} {\bibfield  {journal} {\bibinfo
  {journal} {Phys. Rev. Lett.}\ }\textbf {\bibinfo {volume} {77}},\ \bibinfo
  {pages} {3865} (\bibinfo {year} {1996})}\BibitemShut {NoStop}%
\bibitem [{\citenamefont {Tian}\ \emph {et~al.}(2009)\citenamefont {Tian},
  \citenamefont {Ye},\ and\ \citenamefont {Jin}}]{tian2009proper}%
  \BibitemOpen
  \bibfield  {author} {\bibinfo {author} {\bibfnamefont {Y.}~\bibnamefont
  {Tian}}, \bibinfo {author} {\bibfnamefont {L.}~\bibnamefont {Ye}}, \ and\
  \bibinfo {author} {\bibfnamefont {X.}~\bibnamefont {Jin}},\ }\href@noop {}
  {\bibfield  {journal} {\bibinfo  {journal} {Physical review letters}\
  }\textbf {\bibinfo {volume} {103}},\ \bibinfo {pages} {087206} (\bibinfo
  {year} {2009})}\BibitemShut {NoStop}%
\bibitem [{\citenamefont {von Klitzing}(2017)}]{QHE}%
  \BibitemOpen
  \bibfield  {author} {\bibinfo {author} {\bibfnamefont {K.}~\bibnamefont {von
  Klitzing}},\ }\href {\doibase 10.1146/annurev-conmatphys-031016-025148}
  {\bibfield  {journal} {\bibinfo  {journal} {Annual Review of Condensed Matter
  Physics}\ }\textbf {\bibinfo {volume} {8}},\ \bibinfo {pages} {13} (\bibinfo
  {year} {2017})},\ \Eprint
  {http://arxiv.org/abs/https://doi.org/10.1146/annurev-conmatphys-031016-025148}
  {https://doi.org/10.1146/annurev-conmatphys-031016-025148} \BibitemShut
  {NoStop}%
\bibitem [{\citenamefont {Miyasato}\ \emph {et~al.}(2007)\citenamefont
  {Miyasato}, \citenamefont {Abe}, \citenamefont {Fujii}, \citenamefont
  {Asamitsu}, \citenamefont {Onoda}, \citenamefont {Onose}, \citenamefont
  {Nagaosa},\ and\ \citenamefont {Tokura}}]{miyasato2007crossover}%
  \BibitemOpen
  \bibfield  {author} {\bibinfo {author} {\bibfnamefont {T.}~\bibnamefont
  {Miyasato}}, \bibinfo {author} {\bibfnamefont {N.}~\bibnamefont {Abe}},
  \bibinfo {author} {\bibfnamefont {T.}~\bibnamefont {Fujii}}, \bibinfo
  {author} {\bibfnamefont {A.}~\bibnamefont {Asamitsu}}, \bibinfo {author}
  {\bibfnamefont {S.}~\bibnamefont {Onoda}}, \bibinfo {author} {\bibfnamefont
  {Y.}~\bibnamefont {Onose}}, \bibinfo {author} {\bibfnamefont
  {N.}~\bibnamefont {Nagaosa}}, \ and\ \bibinfo {author} {\bibfnamefont
  {Y.}~\bibnamefont {Tokura}},\ }\href@noop {} {\bibfield  {journal} {\bibinfo
  {journal} {Physical review letters}\ }\textbf {\bibinfo {volume} {99}},\
  \bibinfo {pages} {086602} (\bibinfo {year} {2007})}\BibitemShut {NoStop}%
\bibitem [{\citenamefont {Venturini}\ \emph {et~al.}(1991)\citenamefont
  {Venturini}, \citenamefont {El~Idrissi},\ and\ \citenamefont
  {Malaman}}]{venturini1991magnetic}%
  \BibitemOpen
  \bibfield  {author} {\bibinfo {author} {\bibfnamefont {G.}~\bibnamefont
  {Venturini}}, \bibinfo {author} {\bibfnamefont {B.~C.}\ \bibnamefont
  {El~Idrissi}}, \ and\ \bibinfo {author} {\bibfnamefont {B.}~\bibnamefont
  {Malaman}},\ }\href@noop {} {\bibfield  {journal} {\bibinfo  {journal}
  {Journal of magnetism and magnetic materials}\ }\textbf {\bibinfo {volume}
  {94}},\ \bibinfo {pages} {35} (\bibinfo {year} {1991})}\BibitemShut {NoStop}%
\bibitem [{\citenamefont {Venturini}\ \emph {et~al.}(1996)\citenamefont
  {Venturini}, \citenamefont {Fruchart},\ and\ \citenamefont
  {Malaman}}]{venturini1996incommensurate}%
  \BibitemOpen
  \bibfield  {author} {\bibinfo {author} {\bibfnamefont {G.}~\bibnamefont
  {Venturini}}, \bibinfo {author} {\bibfnamefont {D.}~\bibnamefont {Fruchart}},
  \ and\ \bibinfo {author} {\bibfnamefont {B.}~\bibnamefont {Malaman}},\
  }\href@noop {} {\bibfield  {journal} {\bibinfo  {journal} {Journal of alloys
  and compounds}\ }\textbf {\bibinfo {volume} {236}},\ \bibinfo {pages} {102}
  (\bibinfo {year} {1996})}\BibitemShut {NoStop}%
\bibitem [{\citenamefont {Wang}\ \emph {et~al.}(2019)\citenamefont {Wang},
  \citenamefont {Yin}, \citenamefont {Fujitsu}, \citenamefont {Hosono},\ and\
  \citenamefont {Lei}}]{wang2019near}%
  \BibitemOpen
  \bibfield  {author} {\bibinfo {author} {\bibfnamefont {Q.}~\bibnamefont
  {Wang}}, \bibinfo {author} {\bibfnamefont {Q.}~\bibnamefont {Yin}}, \bibinfo
  {author} {\bibfnamefont {S.}~\bibnamefont {Fujitsu}}, \bibinfo {author}
  {\bibfnamefont {H.}~\bibnamefont {Hosono}}, \ and\ \bibinfo {author}
  {\bibfnamefont {H.}~\bibnamefont {Lei}},\ }\href@noop {} {\bibfield
  {journal} {\bibinfo  {journal} {arXiv preprint arXiv:1906.07986}\ } (\bibinfo
  {year} {2019})}\BibitemShut {NoStop}%
\end{thebibliography}%

\end{document}